\documentclass[12pt,a4paper]{article}      

\usepackage{wrapfig}
\usepackage{graphicx}
\usepackage{epsfig}

\begin{document}

\def\figurename{Fig.}

\begin{center}
	{\Large \bf Production of charmed particles in proton-proton and light  nucleus-nucleus interactions in Geant4 FTF model } 
\end{center}

\begin{center}
{\bf            
A. Galoyan$^{1}$,
A. Ribon$^{2}$,
V. Uzhinsky$^{3}$\\
}
\vspace{0.5cm}
on behalf of the Geant4 hadronic physics working group
\end{center}

\begin{center}
    \begin{minipage}{11cm}
$^{1}$VBLHEP, JINR, Dubna, Moscow region, 141980 Russia\\
$^{2}$CERN, CH-1211 Genève 23, Geneva, Switzerland\\
$^{3}$MLIT, JINR, Dubna, Moscow region, 141980 Russia
    \end{minipage}
\end{center}

\begin{center}
	\begin{minipage}{12cm}
The total yield of neutral D-mesons  (4$\pi$) in central {\rm Xe}+{\rm La} interactions at 150 GeV per nucleon 
has been calculated within  the Geant4 FTF model. Our calculation is close to predictions 
of Monte Carlo models that do not account for quark-gluon plasma formation.  As the predictions, our calculation  
substantially underestimates preliminary experimental data. The experimental data indicate enhanced charmed particle production in nucleus-nucleus interactions. We present also our calculations for charmed meson production in
${\rm p+p}$, ${\rm d+d}$ and ${\rm ^4He+^4He}$ interactions for future NICA/SPD experiment at $\sqrt{s_{NN}}=$10
and 20 GeV.  
    \end{minipage}
\end{center}


\label{sec:intro}
\section*{Introduction}

Recently, the NA61/SHINE collaboration has released preliminary experimental data on the production of $D^0$ and 
anti-$ D^0$  mesons in central (0--20 \%) {\rm Xe}+{\rm La} interactions at 150 GeV/nucleon  \cite{Merzlaya:2024cbt}. 
These data were compared with predictions of several widely used Monte Carlo models --PYTHIA+Angantyr, PHSD, HSD, 
and AMPT. They are presented  in Fig. 1 (see references to the models and the explanation of the figure in
 \cite{Merzlaya:2024cbt}). It turned out that the models significantly underestimate the yield of neutral $D$ mesons. 
Most likely, not only the well-known enhanced strange particle production but also the enhanced yield of charmed particles in nucleus-nucleus collisions occurs.

At the same time, predictions of ALCOR and SMES models overestimate the data  \cite{Merzlaya:2024cbt}. 
These models effectively take into account effects of the quark-gluon plasma (QGP) formation.

We believe that the implementation of Spin Physics Detector (SPD) experiment
\cite{SPDproto:2021hnm,Abramov:2021vtu,SPD:2024gkq} at the NICA accelerator complex will help address the 
issues of models. The main goal of the experiment is to study the gluon composition of polarized and unpolarized 
 light nuclei. The aim of our paper is to estimate inclusive cross sections of $D$-mesons produced in  
proton-proton and light nucleus-nucleus interactions in SPD energy range. 

It is commonly assumed that QCD processes involving gluons are responsible for the production of 
 charmed particles.
Although in recent papers by S. Ostapchenko \cite{Ostapchenko:2025jlz, Garzelli:2023jlq} it was necessary 
to take into account the intrinsic charm of nucleons in an approach similar to that of Brodsky et al. \cite{Brodsky:1980pb,Brodsky:1981se} 
in order to provide a good description of LHCb data \cite{LHCb:2016ikn,LHCb:2013xam,LHCb:2015swx} on pp 
interactions at CMS energies of 5, 7, and 13 TeV, in addition to QCD processes. In the present paper, we consider
 the production of charmed particles taking into account only the Schwinger's mechanism 
\cite{Schwinger:1962tp} of the formation of $c\bar c$ quark pairs during string fragmentation within the 
FTF\footnote{FTF is a Geant4 implementation of the well-known FRITIOF model \cite{Andersson:1986gw, 
Nilsson-Almqvist:1986ast}.} generator of the Geant4 package \cite{GEANT4:2002zbu,Allison:2006ve,Allison:2016ll}. 
All our simulations were performed with the public version 11.1 of Geant4.

\begin{figure}[bth]
\begin{center}
\includegraphics[width=150mm]{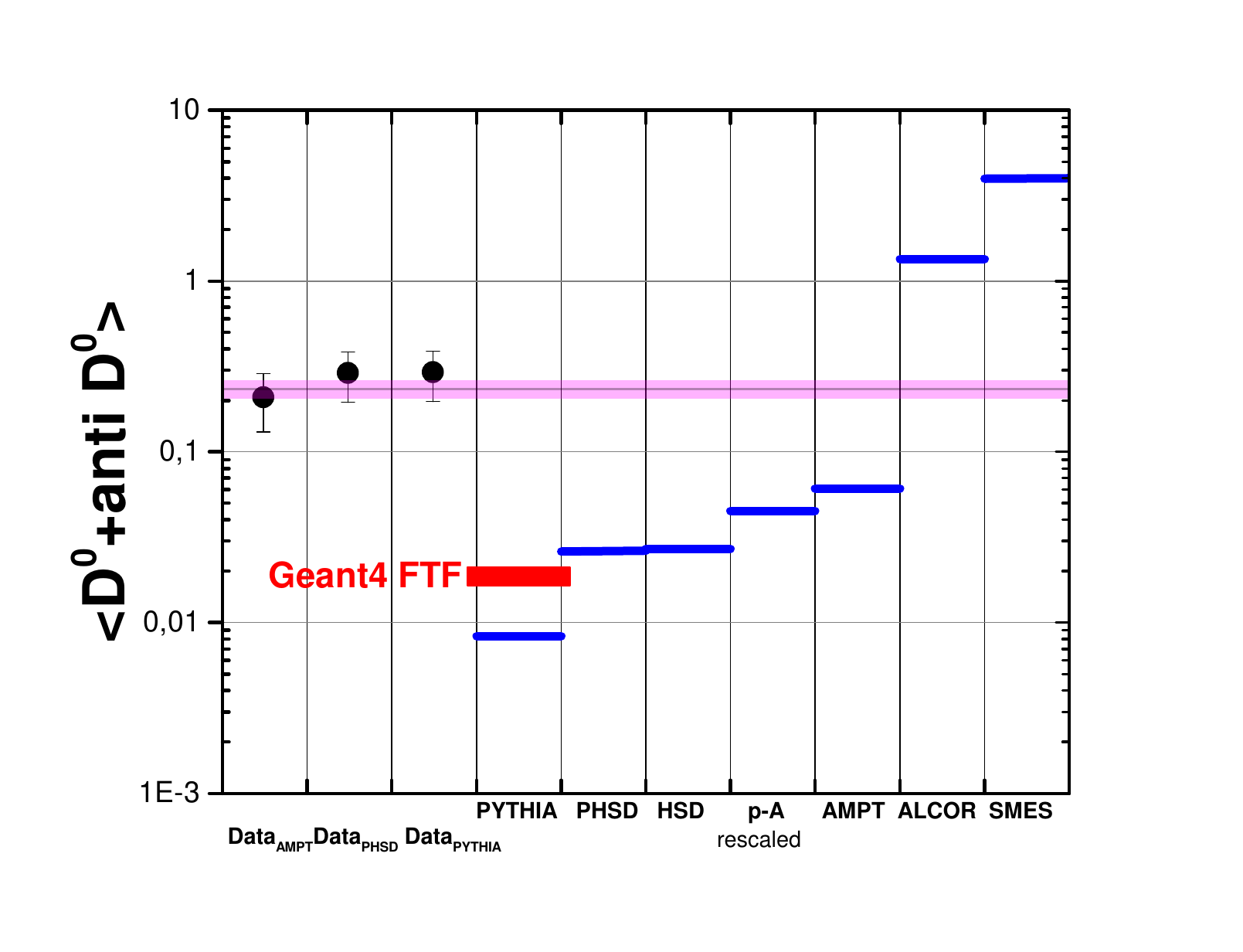}   
\vspace{-3mm}
\caption{Preliminary results of the NA61/SHINE collaboration on the 4$\pi$-yield of $D^0$ + $\bar D^0$ mesons 
compared with model predictions. The horizontal shaded band represents the uncertainty of the experimental 
results due to the unknown phase space distribution of $D^0$  and $\bar D^0$. The points with error bars are the
 experimental data. The horizontal line segments show the model predictions calculated by the collaboration. 
 The red box is the Geant4 FTF model result. "The background in the Monte Carlo (MC) event was
described using the EPOS model, while the signal phase space was parametrized using 3 models: AMPT, PHSD 
and PYTHIA/Angantyr, which predict quite different phase space distributions of open charm." 
\protect{\cite{Merzlaya:2024cbt}} Thus, there are three experimental points.}
\end{center}
\label{fig01}
\vspace{-5mm}
\end{figure}

\label{sec:Geant4Models}
\section{Geant4 High Energy Models}
There are two models in Geant4 for simulating of charmed particle production: FTF and  QGS (Quark-Gluon String)
models \cite{GaloyanCharm}. Only FTF can be applied for nucleus-nucleus interactions. The models suggest various mechanisms for the formation 
of quark strings. The FRITIOF (FTF) model \cite{Andersson:1986gw,Nilsson-Almqvist:1986ast} assumes that two strings 
are produced in inelastic, non-diffractive hadron-nucleon collisions. One string appears  in diffractive interactions. 
The parameterizations of the cross sections of the corresponding processes were determined in Geant4 by analyzing 
a large body of experimental data. A detailed description of the FTF and QGS models can be found in Geant4 Physics 
Reference Manual
https://geant4-userdoc.web.cern.ch/UsersGuides /PhysicsReferenceManual/html/index.html.

The QGS model was proposed in papers by A.B. Kaidalov and K.A. Ter-Martirosyan \cite{Kaidalov:1982xg,Kaidalov:1982xe}.
It was further developed in \cite{Kaidalov:1985jg,Kaidalov:1986zs,Kaidalov:1986zf}. In general, it is based on the Regge phenomenology. The cross sections for string formation and diffraction processes are determined using
 the quasi-eikonal Pomeron approach. 
 
When simulating string fragmentations, an iterative algorithm based on the paper by R. Feynman and 
E. Field \cite{Field:1976ve} is used.  It is assumed that a quark, or an antiquark, or a di-quark located at the end of a string
pick up an antiquark, or a quark from Schwinger's pair produced in the string color field;  and passes into a hadron and 
a residual quark system. The vertices of the most important processes of charmed particle formation are shown 
in Fig. 2. Fragmentation functions are used to determine the kinematic characteristics of hadrons. 
They are distributions of hadrons on transverse momentum ($p_T$) and the scaling variable $x$, defined, 
in the simplest case,  as a ratio of the longitudinal momentum of the hadron  to longitudinal momentum of the quark.

\begin{figure}[bth]
\begin{center}
\includegraphics[width=127mm]{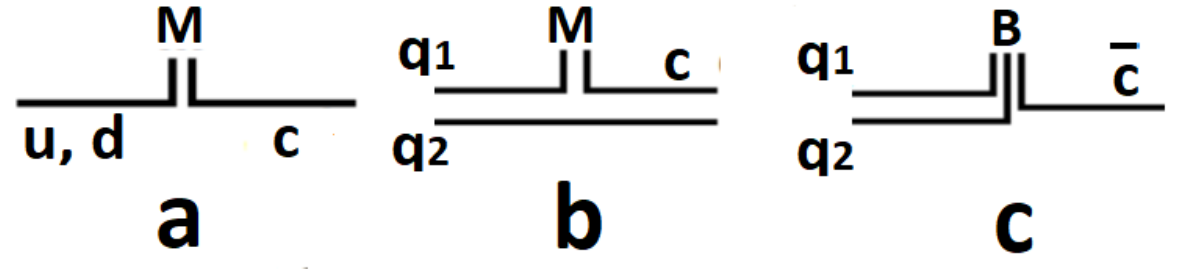}
\vspace{-3mm}
\caption{Processes of quark and di-quark fragmentations into a charmed meson (a, b) , and fragmentation of 
a di-quark into a charmed baryon (c).}
\end{center}
\label{fig02}
\vspace{-5mm}
\end{figure}

The FTF model employs the Lund fragmentation function \cite{Andersson:1983ia}, while the QGS model utilizes
 the functions proposed in \cite{Kaidalov:1985jg}. Transverse momenta of hadrons, $p_T$, are sampled under 
the assumption of  the $m_T$ scaling. The FTF model results are presented in Figs. 3, 4, and 5.  The results of 
the QGS model for $pp$ interactions are close to the FTF ones.

The most important  parameter in both  the models is the probability of producing an anticharm-charm pair
($c\bar c$)  in the color-string field, denoted  as $P_{c\bar c}$. To determine  this probability, we compared 
the model predictions with  the experimental data compiled in the review paper \cite{Lourenco} (see Fig. 3). 
The optimal value was found to be $P_{c\bar c}=$0.0002, which is in good agreement  with the  value reported 
in \cite{Piskunova:1985sp}.

\begin{figure}[bth]
\begin{center}
\includegraphics[width=140mm]{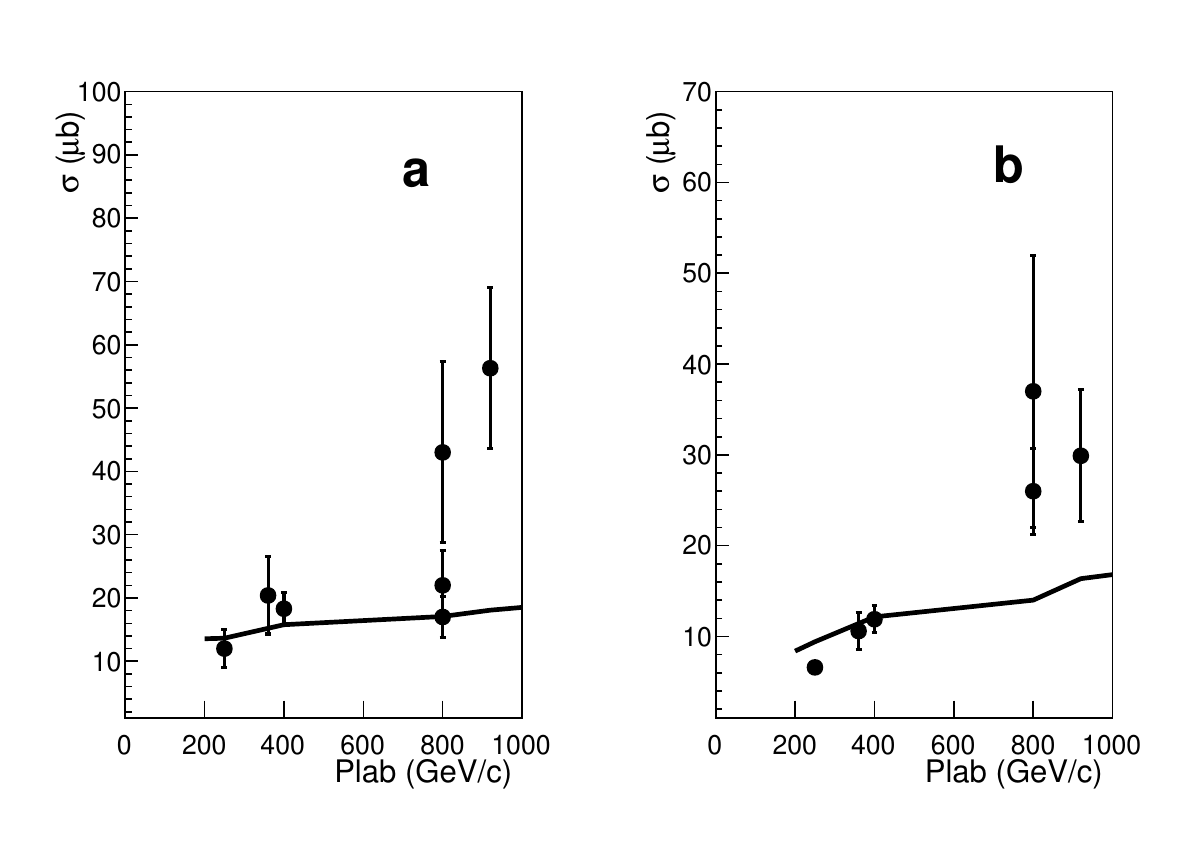}         
\vspace{-3mm}
\caption{Total inclusive cross sections of $D^0$ and $\bar D^0$ (a) and $D^+$ and $D^-$ (b) in $pp$ interactions as functions of projectile momenta in the laboratory system. The points are experimental data collected in
	\protect{\cite{Lourenco}}.  The curves are our FTF model calculations.}
\end{center}
\label{fig03}
\vspace{-5mm}
\end{figure}

Various $D$ mesons distributions depending of  the Feynman variable $x_F$ are well described at  the chosen parameter $P_{c\bar c}$ (see Fig. 4). To describe the distributions over the square of the transverse
 momentum, $p^2_T$, we varied the effective “temperature” of the $m_T$ spectra and adopted $T =$ 200 MeV. 
As seen from Fig. 4, we achieved  a satisfactory description of the experimental data.

\begin{figure}[bth]
\begin{center}
\includegraphics[width=115mm]{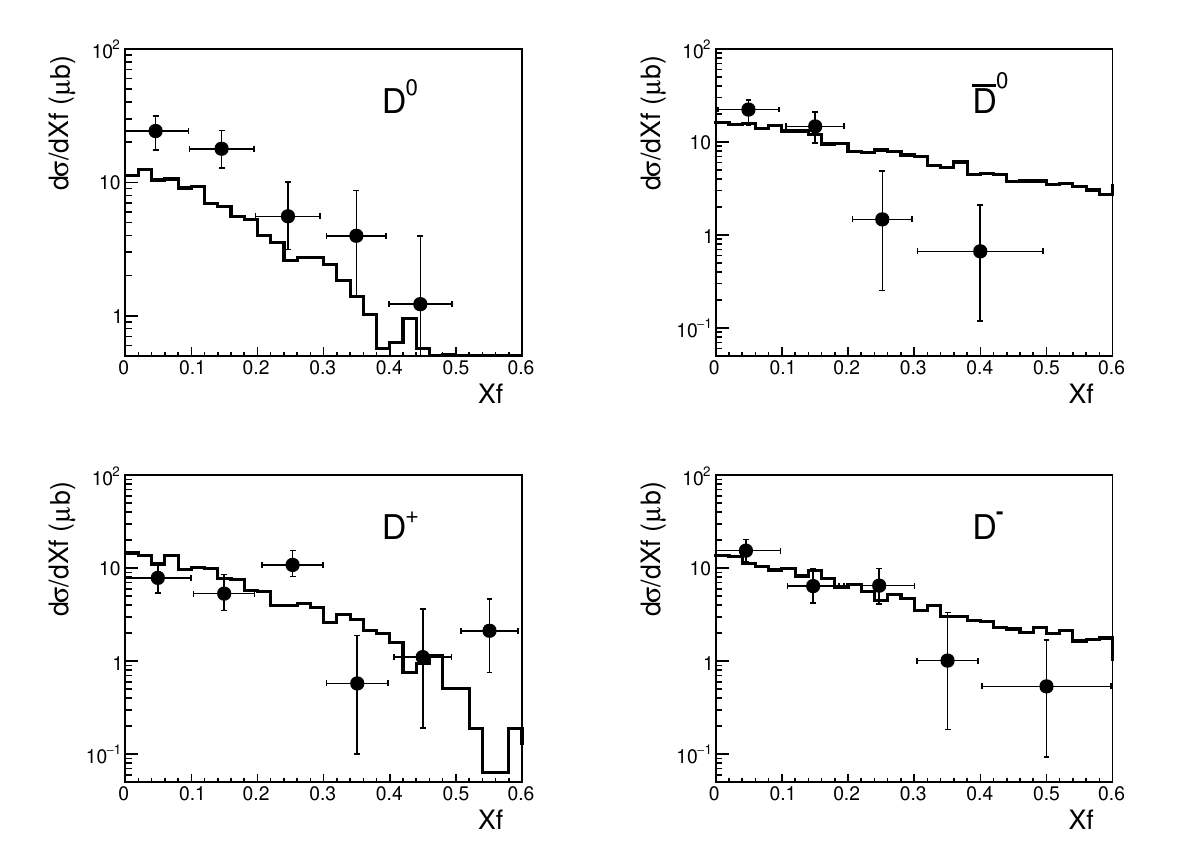}   
\includegraphics[width=115mm]{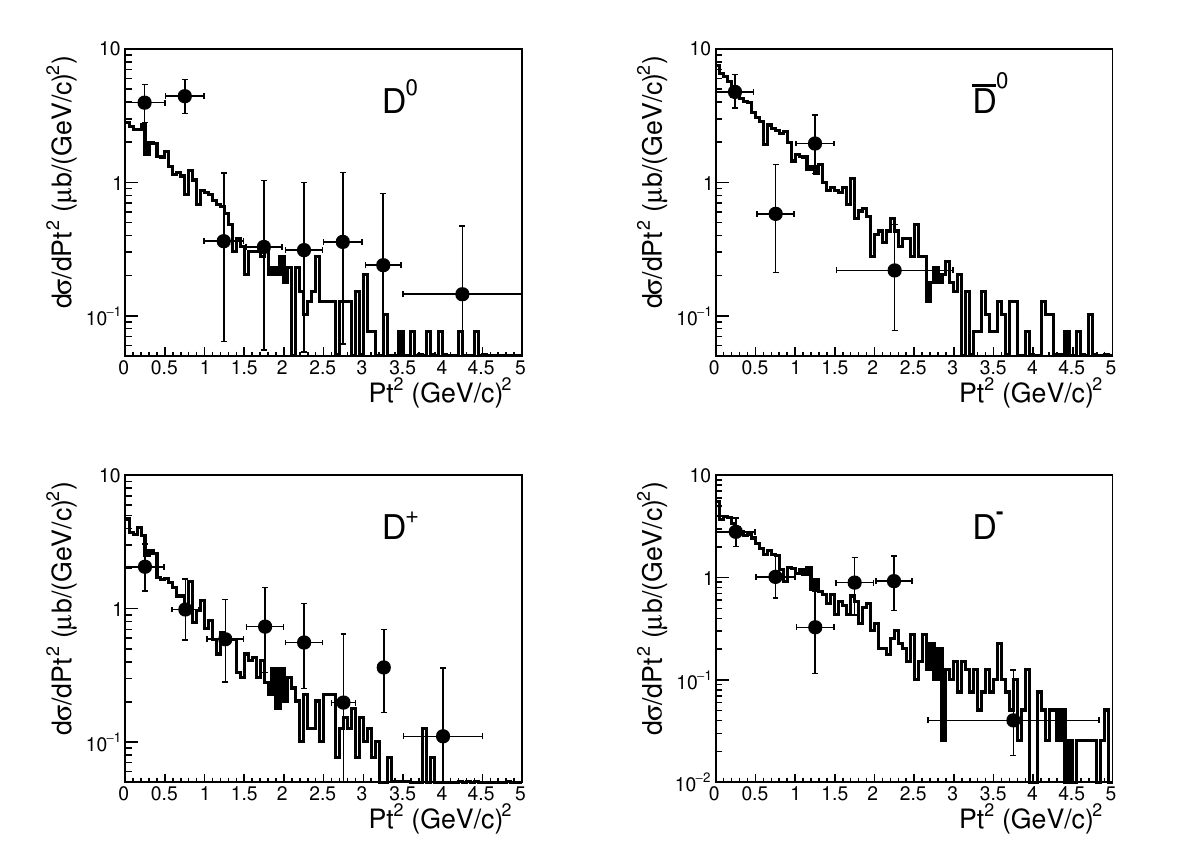} 
\vspace{-3mm}
\caption{Properties of charmed hadrons produced in $pp$ interactions at 400 GeV/c. The points are LEBC-EHS 
collaboration data \protect\cite{LEBC-EHS:1988oic}. Histograms  show our calculations with the FTF model. 
The QGS model gives analogous results. }
\end{center}
\label{fig04}
\vspace{-5mm}

\end{figure}

In Fig. 5, we present our FTF model predictions for total yields of $D^0$, $\bar D^0$, $D^+$ and $D^-$ mesons 
produced  in ${\rm p+p}$, ${\rm d+d}$ and $^4{\rm He}+^4{\rm He}$ interactions for energies achievable in the future SPD 
experiment. As seen from Fig. 5, we obtained that the multiplicity of all D-meson species at $\sqrt{s}$=10 GeV 
is lower than one at  $\sqrt{s}$=20  GeV.  The differential cross sections of D-mesons in ${\rm He+He}$
interactions  are larger than those in ${\rm d+d}$ interactions. The differential cross sections  of D-mesons in  
${\rm p+p}$ interactions are smaller than ones in ${\rm d+d}$ collisions. Although the shapes of the distributions are 
essentially the same for the different interactions. They differ only in overall normalization, which scales with 
the average number of inelastic collisions in nucleus-nucleus interactions. For heavier colliding nuclei and at higher energies, 
one can anticipate the onset of hard process effects and the possible formation of a quark-gluon plasma.

\begin{figure}[bth]
\begin{center}
\includegraphics[width=110mm]{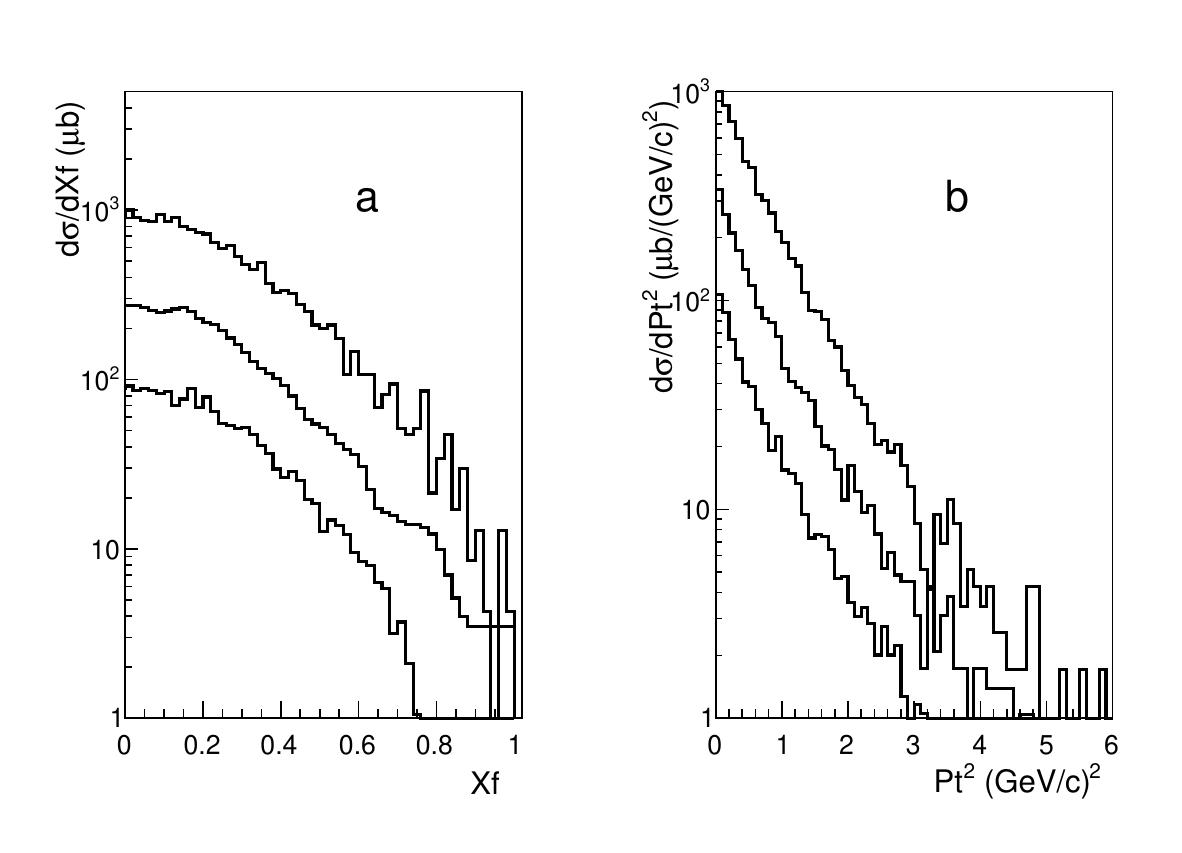} 
\includegraphics[width=110mm]{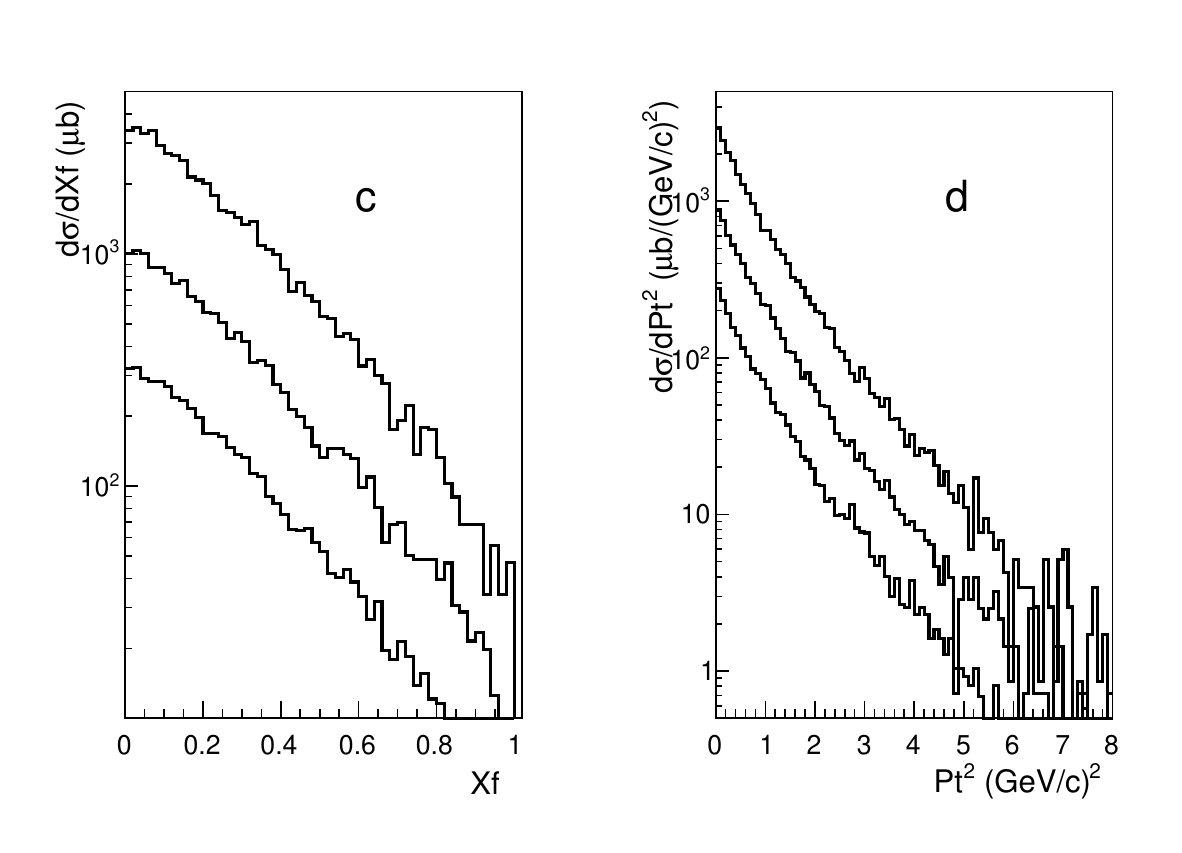}

\vspace{-3mm}
\caption{Inclusive distributions of $D$ meson ($D^0$, $\bar D^0$, $D^+$ and $D^-$) in ${\rm p+p}$, ${\rm d+d}$ and
 $^4{\rm He}+^4{\rm He}$ interactions. The upper histograms correspond to $^4{\rm He}+^4{\rm He}$ interactions. 
The low histogram are for ${\rm p+p}$ ones. The middle histograms are for ${\rm d+d}$ interactions. The upper figures (a, b) are for $\sqrt{s}=10$ GeV, the low figures (c, d) are  for $\sqrt{s}=20$ GeV }. 
\end{center}
\label{fig05}
\vspace{-5mm}
\end{figure}

Our results, presented in Fig. 3 and 4, show that we describe the experimental data on hadron-nucleon interactions quite well. Therefore, we have confidence in our predictions for pp collisions (Fig. 5). The probability of multi-nucleon interactions 
in ${\rm dd}$ collisions is $\sim 30$\%. The same probability in the case of ${\rm He+He}$ is $\sim 46$\%. 
Perhaps, the contributions of the multi-nucleon interactions will be larger than the FTF model predictions.
We hope that a detailed study of the charmed particle production in light nuclei interactions at SPD will clarify  the role of 
the multi-nucleon collisions in the charmed quark creation. 


Our calculation using the FTF model for {\rm Xe+La} interactions at 150 GeV per nucleon is shown in Fig. 1. It is close to 
the predictions of models that do not incorporate quark–gluon plasma formation. Other mechanisms for 
charm meson production are, of course, possible and still need to be identified.

\section*{Conclusions}
\begin{enumerate}
\item The existing Geant4 implementation of charmed meson production in hadron-nucleon interactions provides  
a satisfactory description of the available experimental data for hadron-nucleus interactions in the SPS energy range.

\item We applied the Geant4 FTF model to describe the preliminary data of the NA61/SHINE collaboration 
on neutral charmed  $D$-meson production in {\rm  Xe+La} interactions. Our resulting yield is comparable to those
obtained with models that do not incorporate quark-gluon plasma formation.

\item The NA61/SHINE experimental data indicate enhanced charmed hadron production in nucleus-nucleus interactions
compared with the predictions of models without QGP.

\end{enumerate}

The authors of the paper are grateful to the team of the JINR HybriLit cluster  for support of our calculations. 
V. Uzhinsky is thankful to the grant of the Government of the Russian
 Federation (Agreement No. 075-15-2025-009, dated February 28, 2025) for partial support of the study.

The authors of this work declare that they have no conflicts of interest.

\def\bibname{References}  

\end{document}